# Observation of Conventional Near Room Temperature Superconductivity in Carbonaceous Sulfur Hydride

Hiranya Pasan[1], Elliot Snider[2], Sasanka Munasinghe[2], Sachith E. Dissanayake[2], Nilesh P. Salke[3], Muhtar Ahart[3], Nugzari Khalvashi-Sutter[2], Nathan Dasenbrock-Gammon[1], Raymond McBride[2], G. Alexander Smith[4,5], Faraz Mostafaeipour[4,6], Dean Smith[4], Sergio Villa Cortés[4], Yuming Xiao[7], Curtis Kenney-Benson[7], Changyong Park[7], Vitali Prakapenka[8], Stella Chariton[8], Keith V. Lawler[4], Maddury Somayazulu[7,*], Zhenxian Liu[3], Russell J. Hemley[3,9,10,*], Ashkan Salamat[4,6,*], Ranga P. Dias[1,2,*]

[1]Department of Physics and Astronomy, University of Rochester, Rochester, NY 14627, USA.
[2]Department of Mechanical Engineering, School of Engineering and Applied Sciences, University of Rochester, Rochester, NY 14627, USA.
[3]Department of Physics, University of Illinois Chicago, Chicago, Illinois 60607, USA
[4]Nevada Extreme Conditions Laboratory, University of Nevada, Las Vegas, Nevada 89154, USA
[5]Department of Chemistry and Biochemistry & Astronomy, University of Nevada Las Vegas, Las Vegas, Nevada 89154, USA
[6]Department of Physics & Astronomy, University of Nevada, Las Vegas, Las Vegas, Nevada 89154, USA
[7]HPCAT, X-ray Science Division, Argonne National Laboratory, Argonne, Illinois 60439, USA
[8]Center for Advanced Radiation Sources, University of Chicago, Chicago, Illinois 60439, USA
[9]Department of Chemistry, University of Illinois Chicago, Chicago, Illinois 60607, USA
[10]Department of Earth and Environmental Sciences, University of Illinois Chicago, Chicago, Illinois 60607, USA

**The phenomenon of high temperature superconductivity, approaching room temperature, has been realized in a number of hydrogen-dominant alloy systems under high pressure conditions[1-12]. A significant discovery in reaching room temperature superconductivity is the photo-induced reaction of sulfur, hydrogen, and carbon that initially forms as van der Waals solids at sub-megabar pressures. Carbonaceous sulfur hydride has been demonstrated to be tunable with respect to carbon content, leading to different superconducting final states with different structural symmetries. A modulated AC susceptibility technique adapted for a diamond anvil cell confirms a $T_c$ of 260 kelvin at 133 GPa in carbonaceous sulfur hydride. Furthermore, direct synchrotron infrared reflectivity measurements on the same sample under the same conditions reveal a superconducting gap of ~85 meV at 100 K in close agreement with the expected value from Bardeen-Cooper-Schrieffer (BCS) theory[13-18]. Additionally, x-ray diffraction in tandem with AC magnetic susceptibility measurements above and below the superconducting transition temperature, and as a function of pressure at 107-133 GPa, reveal the *Pnma* structure of the material is responsible for the close to room-temperature superconductivity at these pressures.**

## Main

The determination of the underlying interaction that governs the formation of Cooper pairs has become one of the major goals in the understanding of new superconducting materials[17-21]. In particular, hydrogen-dominant alloys, such as the superhydrides, have been found to exhibit high superconducting critical temperatures ($T_c$) in excess of 200 K at megabar pressures (>100 GPa)[1-8,10-12]. These materials, including both binary and an increasing number of ternary hydride systems, have been explored by simulation — with the theoretical calculations in some cases first predicting in advance high $T_c$ superconductivity in the phases subsequently found by experiment[2,22-30]. These calculations provide insight as to how electron-phonon interactions give rise to the formation of Cooper pairs and phonon mediated superconductivity[17,18,31-36]. High-$T_c$ superconductivity arises in these materials as a result of a combination of a high density of states near the Fermi level, high-frequency phonon modes, and strong electron-phonon interactions[19,20,37]. Within BCS theory, the Cooper pair binding energy, which is the difference between the normal and superconducting states, is known as the superconducting gap. The superconducting gap in the zero temperature limit is $2\Delta_0 \sim 3.5 k_B T_C$, and is on the order of meV[13-16]. Detailed measurements of superconducting properties such as the coherence length, penetration depth, and superconducting energy gap under pressure, particularly in the megabar regime required for these hydrogen-rich materials, present numerous challenges in comparison to ambient pressures.

For a conventional phonon-mediated superconductor, the essential parameters that determine the thermodynamic and electromagnetic properties are the electron-phonon spectral density ($\alpha^2 F(\omega)$; the Eliashberg function) and the Coulomb repulsion parameter ($\mu^*$)[31,34,38]. The Eliashberg function can be extracted from tunneling data for many conventional superconductors and used, with considerable success, to compute their properties[31]. Due to experimental challenges at the extremely high pressures required to study the new hydrogen-rich superconductors, tunneling experiments cannot be done. Alternatively, optical measurements at long wavelengths can in principle be used to probe characteristic features of the superconducting state[31]. This is demonstrated at ambient pressure for numerous conventional superconductors, i.e., those with relatively low critical temperatures for which the gap is in the far-IR range. However, optical measurements of the superconducting gap can be limited by the small sample size for which the energy of the gap is below the diffraction limit of the light. For such samples under very high-pressure conditions, multiple probes are needed to unambiguously characterize the superconducting state of these materials. For example, x-ray diffraction measurements of structural properties of the superconducting phase was shown to be essential in the work leading to the discovery of very high $T_c$ superconductivity in LaH$_{10}$[2]. However, x-ray diffraction of the material in the superconducting state has not been measured for other superconductors, including both low- and high $T_c$ materials, at these high pressures.

Here we report the application of multiple measurements on the synthesized carbonaceous sulfur hydride (C-S-H) superconductor with very high $T_c$ at lower pressure, and using new higher sensitivity techniques. We use our previously described method of compositional tuning of $T_c$ for the synthesis of the material,[9] and apply a modulated AC susceptibility technique for identifying the onset of superconductivity. The results reveal a superconducting transition of ~260 K at 133 GPa for a C-S-H sample, which is a significantly lower pressure for reaching the same $T_c$ compared to that found in the original syntheses. Synchrotron IR measurements from 70 to 170 meV on the same sample reveal a sharp decrease in reflectivity compared to the normal state at temperatures below $T_c$, and give a superconducting gap of ~85 meV at 100 K. The measured gap closely agrees with the BCS value of ~80 meV for a $T_c$ of 260 K, indicating the conventional nature of the superconductivity. Finally, the structure of the superconducting phase was directly probed by synchrotron single-crystal x-ray diffraction whilst simultaneously measuring the Meissner effect using AC susceptibility. In addition to confirming the very high $T_c$ previously found in the C-S-H system at higher pressures, this work suggests new pathways for realizing room-temperature superconductivity at still lower pressures in this system as well as providing new information on the structure, optical, and electronic properties important for building reliable theoretical models for these novel high $T_c$ superconducting phases.

## Magnetic Susceptibility Measurements

Direct measurement of the magnetic behavior of a material on passage into the superconducting state, and specifically the expulsion of external magnetic fields, is essential to confirm superconductivity. This characteristic is referred to as the Meissner effect and within the confinements of the DAC is measured using magnetic susceptibility techniques. A two-coil setup is often utilized, in which an AC magnetic field is applied for one of the coils and the other is used to detect the field by measuring the induced voltage[39]. The coil generating the field is called the "primary/excitation coil" and the coil detecting the field is called the "secondary/pickup coil". Since the sample is placed in the middle of the coil, the induced electromotive force (voltage) is proportional to the average magnetic susceptibility of the volume contained within the coil, which is proportional to the sample's magnetic susceptibility if other conditions remain the same. This basic method is known as the signal-frequency AC susceptibility method. However, in practice, variable environmental conditions give rise to a residual background voltage due to geometrical differences between coils, asymmetrical deformation of the coils with regards to temperature, along with the non-uniform magnetic flux due to all the metallic components of the DAC. Such complexities can result in masking the superconducting transition by the non-zero, temperature-dependent background. To enhance the sensitivity of the technique for DAC-sized samples, a double-frequency method was introduced in which the superconducting state at the onset of the transition is suppressed by use of another high-amplitude modulation field[39]. This technique relies on the critical field behavior of a superconductor. The excitation field is much smaller and at higher frequency than the modulation field and measures the diamagnetic change in the sample. The modulation field is thus at a lower frequency and higher strength than the excitation field. For a small region in temperature, the modulation field exceeds the critical field. This results in the superconducting state being suppressed at twice the modulation frequency, since this suppression will occur at both the peak and the trough of the modulation field.

The decreasing size of samples, including small single crystals, of the novel superconductors at megabar pressures has required the developments of ever more sensitive magnetic susceptibility techniques. In the present work, we present the results of a significantly improved AC magnetic susceptibility technique (see Methods Section). Samples were photochemically synthesized at ~4 GPa following the same protocol as Snider et al. and Smith et al. and then compressed to >100 GPa for all subsequent measurements (Fig. 1a)[4,9,40]. A schematic of the improved technique is shown in Fig. 1b,c. Selected results obtained with the improved instrumentation are shown in Fig. 1d-f. The superconducting transition is apparent from the abrupt change in signal at the onset of $T_c$. The $T_c$ corresponds to the onset of the susceptibility signal on the high-temperature side, where the magnetic flux completely enters the sample. The onset of superconductivity is identified by the appearance of the asymmetric peak indicating a diamagnetic transition, which shifts to higher temperatures with increasing pressure. Effects of non-hydrostaticity skew the peak shape, causing a more Gaussian-like profile. The highest transition temperature measured in this way was 260 K (blue arrow Fig. 1f) reached at the highest pressure measured (133 GPa).

The pressure dependence of superconducting transition temperature of a C-S-H sample measured with this technique between 107 and 133 GPa is shown in Fig. 2. The $T_c$ monotonically increases from 202 to 260 K across the measured pressure range. Thus, the extrema in $T_c$ denoting a possible peak of the superconducting dome for this phase of C-S-H was not observed. The superconducting response measured here presents as a continuation of that measured in Smith et al. where a max $T_c$ of 191 K at 97 GPa was reported (Fig. 2). This is clearly a lower pressure superconducting transition than presented in the original report of C-S-H (see Fig. 2 inset). The key difference between these superconducting responses is that the vibrational Raman spectra of the synthesized materials here and in Smith et al., show a different carbon concentration than that of Snider et al[4,9]. This gives rise to a different phase evolution[41,42], and thus high-$T_c$ superconductivity at lower pressures.

**Synchrotron Infrared Measurements of Superconducting State**

Conventional superconductivity is well explained through the electron-phonon coupling and the formation of Cooper pairs, which was established by the observation of the substantial rise in transmission when the photon energy was decreased toward the gap value from the normal state[43-46]. The experimental measurement of optical conductivity (through far infrared or microwave spectroscopies) enables one to extract the energy gap of single-particle excitations. At $T \ll T_c$ the quasiparticle excitations are frozen out. Since no absorption is possible for $\hbar\omega < 2\Delta_0$, the dissipative part of the complex conductivity, $\sigma_1$, vanishes at all frequencies below the gap but is expected to rise steeply as soon as $\hbar\omega$ exceeds the gap[47]. Optical conductivity experiments cannot be done and instead infrared reflectance measurements at low photon energies provide an effective alternative as finite-frequency absorption onsets at a frequency $\hbar\omega = 2\Delta(T)$ in the superconducting state, where $\Delta(T)$ is the temperature-dependent energy gap. In other words, the reflectivity is unity for frequencies below $2\Delta(T)$, and drops sharply above $2\Delta(T)$[48]. Numerous conventional superconductors probed in this way show a superconducting gap that is well fit by the values predicted with BCS theory[13,15,16,45,46,48-51].

In contrast to these ambient pressure measurements, there have been few direct measurements of superconducting gaps under pressure, in part because the long-wavelength (low photon energy) diffraction limit for conventional IR, THz, and microwave spectroscopy for low-temperature superconductors, where the gap can be below the diffraction limit. Synchrotron infrared spectroscopy combined with diamond anvil cells greatly extended the low-energy limit for broad-band IR measurements and provide important constraints on metallization (and superconductivity) in materials under pressure[52-54].

The C-S-H sample studied by magnetic susceptibility and x-ray diffraction was also examined using synchrotron IR/THz spectroscopy at beamline 22-IR-1, NSLS-II, BNL. The ratio of the reflectivity of the sample in the superconducting state relative to that in the normal state, $R_s/R_n$, is shown in the Figure 3. See methods section for more details. Similar analyses have been carried out for other superconducting materials including $H_3S$[48]. Below 260 K a change in overall shape of the IR reflectance occurs that is associated with transformation to the superconducting state. Figure 3 shows the $R(T)/R_n$ curves for the C-S-H sample at four different temperatures near 133 GPa, where the IR reflectance at 290 K was used for the normal state of the sample. The 260 K curve shows almost no change within 70–170 meV energy range, whereas the 230 K, 150 K and 100 K data reveal a significant decrease in the reflectivity ratio. At 260 K, the sample has not fully reached to its superconducting state, and thus is indicative of more metallic behavior, while the lower temperature data corresponds to the superconducting state of the sample. In addition, temperatures below 260 K, the drop in reflectivity ratio becomes more appreciable and eventually saturates near 100 K.

The result indicates the appearance of the superconducting gap on cooling through the $T_c$ observed by magnetic susceptibility. The maximum superconducting gap is estimated at ~85 meV at 100 K based on the first derivatives of the curves. The results also reveal the enhancement of superconducting behavior with decreasing temperature within 70-170 meV, *i.e.,* in the low energy range. This enhancement saturates at 100 K, and at ~85 meV the superconducting gap feature is observed, close to that predicted by BCS theory for a $T_c$ of 260 K. These IR measurements thus strongly support the conventional electron-phonon coupling is responsible for the very high $T_c$ of the C-S-H superconductor.

## X-ray Diffraction and Structure

Complementing the magnetic susceptibility and synchrotron IR experiments, x-ray diffraction was also measured on the same sample as a function of both pressure and temperature. The structural data match high pressure x-ray patterns reported previously for C-S-H[42,55]. The results show that the *Pnma* phase persists over the pressure range studied. The lattice parameters from different pressures and temperatures are summarized in Table S1. Overall, the new *P-V* data are consistent with previously reported room temperature equation of state results, which show the evolution of C-S-H as well as related precursor phases over a broad pressure range (Fig. 4a)[42,55].

Synthesis and structural determination studies performed subsequent to the initial report of superconductivity in C-S-H confirm the initial low pressure phase is an *I4/mcm* $Al_2Cu$-type structure[9,41]. In two previous studies, a monoclinic *C2/c* phase is seen to emerge in samples with a low carbon concentration around ~20 GPa and persist to ~30 GPa[9,41,42]. Beyond 30 GPa in low carbon concentration samples, the sample can again be indexed as the *I4/mcm* structure[9,42], where Raman measurements indicate this phase is similar to the low pressure *I4/mcm* phase but with ordered molecular subunits[42,55]. In C- (or $CH_4$)-rich systems the monoclinic phase is not observed, and the samples transform directly from the low pressure, disordered to ordered *I4/mcm* phase. Depending on initial reactants and thermodynamic pathway, the ordered *I4/mcm* phase is observed to orthorhombically distort by 124 GPa. This distortion is best refined as *Pnma*, and it is the structure measured here for C-S-H from 107–135 GPa. The *Pnma* structure has a doubled unit cell volume relative to the *I4/mcm* structure, and like the monoclinic phase, it maintains the $Al_2Cu$ packing motif[42,55]. Beyond the *Pnma* phase, an *Im$\bar{3}$m* phase reminiscent of the high-pressure superconducting phase of $H_3S$ has been reported to appear without annealing by 159 GPa, persisting to at least 240 GPa[42].

Previous x-ray diffraction measurements of C-S-H samples which had the same superconducting response as this work indicated the *I4/mcm* phase persists up to 100 GPa,[9] but those measurements were limited to 295 K and did not examine the structure of samples in the superconducting state (*e.g.,* by electrical resistance or magnetic susceptibility). In contrast to previous work, synchrotron x-ray diffraction was measured in tandem with the AC susceptibility and tracked the structure of the material both above and below $T_c$. For all temperatures and pressures measured here, only the orthorhombic *Pnma* structure was observed and again the refinements are in good agreement with previous reports of the phase[9,42]. On the other hand, differences may be expected due to different stoichiometries (*e.g.,* C:S ratio) that give rise to different maximum superconducting temperatures at a given pressure. Diffraction data collected on different instruments reveal that the crystal remains largely single crystal in nature with preferred orientation across the few crystallites. Prominent reflections arising from the (002), (222), (012) and (102) planes are observed. At ~125 GPa, diffraction was measured at 80 K and 295 K, with superconductivity simultaneously confirmed at the lower temperature by AC susceptibility. The integrated intensity versus *d*-spacing plots of major diffraction peaks shows no distortion (Fig. 4b). The movement of peak position to lower *d*-spacing arises from thermal contraction of the lattice, which is about 2% from 295 K to 80 K. Additionally, no new peaks were observed at 80 K, which implies that crystal structure remains same above and below $T_c$. Thus, no observation of a symmetry change or lattice distortion upon passage into the superconducting state is observed.

**Discussion**

Compositional tuning of the carbon content and phase of the C-S-H system has been shown to be a way forward to lower the required pressures for superconductivity while still maintaining very high critical transition temperatures compared to the $H_2S$ and $H_2S+H_2$ derived superconductors. A key result of this multi-technique study is direct x-ray diffraction probe of the structure of the material in the superconducting state, as we show there is no structural symmetry change with the *Pnma* solution above and below the superconducting transition. Similar results were reported for polycrystalline $LaH_{10}$ through its $T_c$ of 260 K[2], but the peaks reported here for C-S-H are clearer because the sample remained largely single-crystalline and the pattern is free of diffraction from electrical leads. The C-S-H diffraction result confirms that the superconducting signal observed by electrical resistivity and magnetic susceptibility is not associated with a major structural change (*i.e.,* within *Pnma*) and thus the $Al_2Cu$ packing motif is maintained in the superconducting state. The question arises of whether specific sublattices of this packing motif are responsible for the superconductivity. Note that diffraction studies of C-S-H to date cannot differentiate crystallographically between the C and S atoms, suggesting that the C and S atoms occupy the same sites[9,41,42,55]. This result has bearing on the possibility of alloying or doping of carbon in the hydrogen sulfide. Notably, hole-doping of cubic (*Im$\bar{3}$m*) $H_3S$ by carbon can explain the enhancement of $T_c$ in C-S-H relative to that of the pure hydrogen sulfide phase (i.e., 203 K to the $T_c$ reported here and previously)[1,56]. Given that the structure of material remains orthorhombic over the conditions studied suggests that the carbon hole-doping model may be applicable to the observed lower symmetry structure. Previously presented DFT simulations with a van der Waals functional of molecular $H_2S$, $CH_4$, and $H_2$ molecular sub-unit packed in the 90 GPa *I4/mcm* structure[9,57] indicated that the $H_2$ molecules are not strongly interacting with themselves or the other sub-lattice. On the other hand, the host "$Al_2$" sub-lattice is evolving from a van der Waals to hydrogen bonding to ice-X like single well interactions between the molecular sub-units where the connectivity of those extended interactions are affected by the orientation and amount of $CH_4$ present. As the structural motif is maintained, it implies that it is likely the evolution of the host sub-lattice with pressure that leads to the observed properties.

Synchrotron IR spectroscopy reveals the superconducting gap of the C-S-H phase within its superconducting state which was monitored by AC susceptibility to have a transition at 260 K at 133 GPa. The results contrast with the infrared measurements of $H_3S$, which focused on the behavior of the mid-IR spectra collected with a conventional IR source[48]. That paper also includes discussion of the low energy range and superconducting gap. Although the use of the synchrotron source was discussed, the data used in the analysis were collected with a conventional source. The measured gap of about ~85 meV at 100 K, is consistent with the BCS theory predicted gap energy of $2\Delta(0) = 3.528 k_B T_C$; with $T_c$ of 260 K independently measured by magnetic susceptibility, the calculated superconducting gap ($2\Delta(0)$) ~80 meV. The good agreement with the BCS prediction supports conventional superconductivity, despite the extraordinarily high $T_c$. In this paper, we study certain samples of C-S-H and establish that our measurement techniques are valid for determining the superconducting transition temperature and validating earlier methods. We have chosen a specific compound of C-S-H that can be produced at lower pressures and focused on the magnetic susceptibility measurements of the Meissner effect. There are also higher pressure structures and compositions of C-S-H, including that measured by Snider at el., which measured the electrical conductivity and magnetic susceptibility, producing a room temperature superconductor.

**Online content**

Extended Data Figures and Supplementary Information linked to the online version of the paper at www.nature.com/nature

**Acknowledgments**

We thank I. F. Silvera and G.W. Collins for useful discussions, A. Lamichhane for help with the x-ray measurements and analysis, and R. C. Heist and L. Koelbl for valuable suggestions on the manuscript. Preparation of diamond surfaces was performed in part at the University of Rochester Integrated Nanosystems Center. This research was supported by the following grants and contracts: National Science Foundation (NSF) grants DMR-1809649 (R.D.) and DMR-2104881 (R.H, N.S.); the Gordon and Betty Moore Foundations EPiQS Initiative, grant GBMF10731 (A.S., R.D.); U.S. Department of Energy (DOE), Office of Science (SC), Fusion Energy Sciences under award DE-SC0020340 (R.D., R.H.); and DOE-National Nuclear Security Administration (NNSA) through the Chicago/DOE Alliance Center (CDAC), cooperative agreement DE-NA0003975 (R.H., M.A., Z.L.); DOE award DE-SC0020303 (A.S., K.L.); COMPRES, the Consortium for Materials Properties Research in Earth Sciences, under cooperative agreement EAR-1606856 (NSF; R.H, Z.L.) and DE-NA0003975 (DOE-NNSA; R.H., Z.L.). Portions of this work were performed at 22-IR-1 beamline of the National Synchrotron Light Source II (NSLS-II), which is a User Facility operated for the DOE SC at Brookhaven National Laboratory under contract DE-AC98-06CH10886; at HPCAT (Sector 16), which is supported by DOE-NNSA Office of Experimental Sciences, and at GSECARS (Sector 13), which is supported by the NSF award EAR-1634415, both at the Advanced Photon Source (APS), Argonne National Laboratory (ANL). The APS is a User Facility operated for the DOE SC by ANL under contract DE-AC02- 06CH11357.

**Author contributions**

H.P., E.S., S.M., and S.E.D contributed equally to this work as co-first authors. H.P., E.S., N.K.S., and G.A.S. synthesized the CSH samples. S.M., S.E.D., N.K.S., and R.P.D., performed the magnetic susceptibility and Raman measurements and analyzed the data. N.D.G. and R.M. assist with the sample preparation and magnetic susceptibility measurements. D. S. analyzed the Raman data. S.E.D., N.P.S., M.A., and S.M. performed the x-ray studies and analyzed the data. A.S., R.J.H., M.S. analyzed the x-ray diffraction results. Y. X., C. K. B., C.P., V.P., and S.C., assisted with the x-ray diffraction measurements and data analysis. H.P., S.E.D., Z.L., and R.P.D. performed the IR reflectivity measurements and analyzed the data, with support of R.J.H., and S.V.C., A.S., and K.V.L. performed the simulations and analyzed the data and chemistry protocol. S.V.C., S.E.D., K.V.L., R.J.H., A.S., R.P.D. wrote the paper. All authors discussed the results and edited the manuscript. R.P.D. conceived of the project and oversaw the entire project.

**Competing interests**

The University of Rochester (U of R) has patents pending related to Ranga Dias' scientific discoveries in the field of superconductivity. Ranga Dias is a Co-founder and Chairman of the Board of Unearthly Materials (UM), Inc., a Delaware corporation. UM has licensing agreements with U of R related to the patents, proprietary interests, and commercialization rights related to Ranga Dias' scientific discoveries. UMI, U of R and Ranga Dias are subject to non-disclosure and confidentiality agreements. Ashkan Salamat is a Co-founder and President and Chief Executive Officer and Board Member of Unearthly Materials, Inc., a Delaware corporation ("UMI").

**Corresponding authors**

Correspondence should be addressed to Ranga P. Dias rdias@rochester.edu, Ashkan Salamat ashkan.salamat@unlv.edu, Russell J. Hemley rhemley@uic.edu, or Maddury Somayazulu zulu@anl.gov.

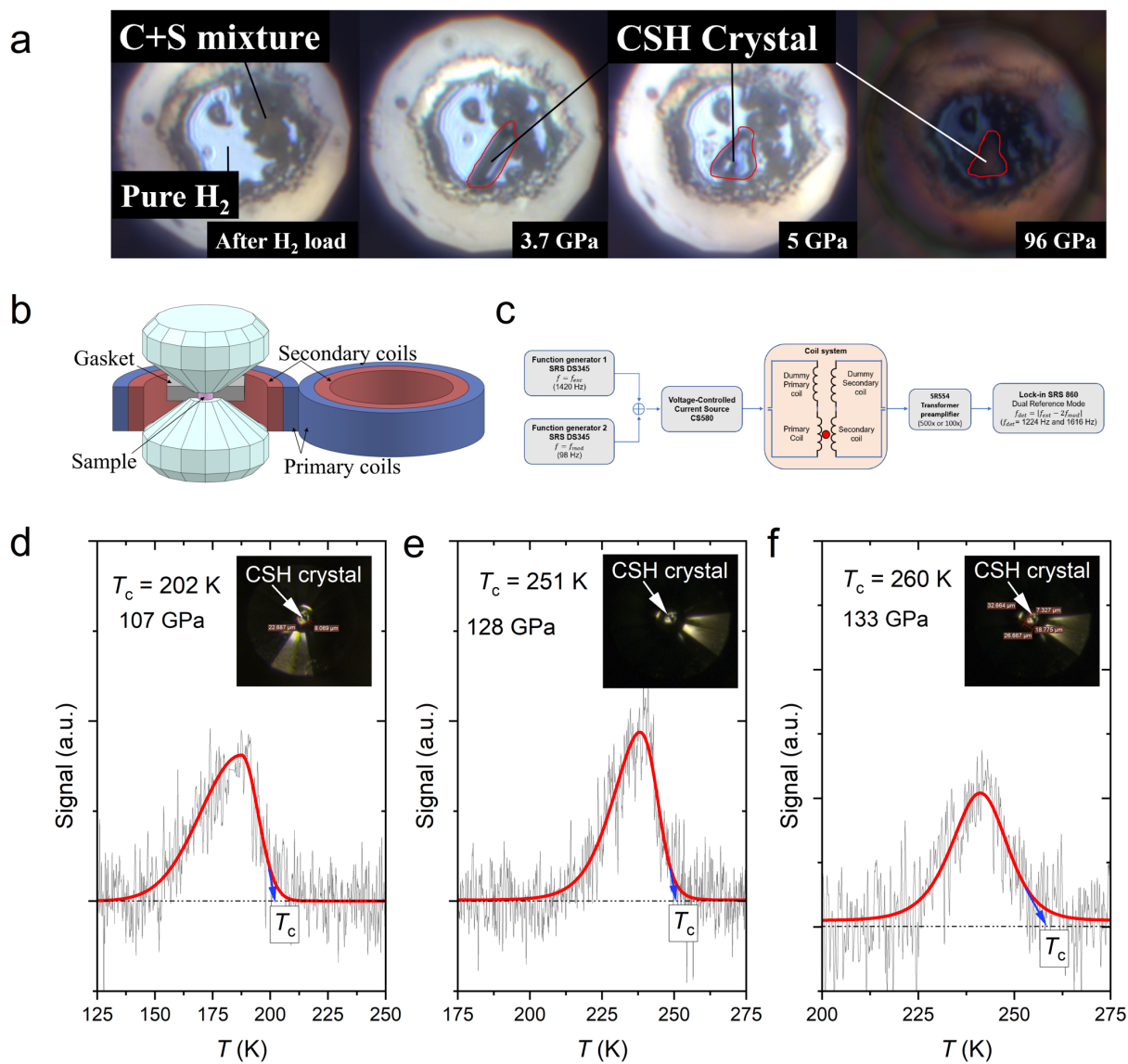

**Figure 1. Modulated AC susceptibility. (a)** Micrographs taken from inside a diamond anvil cell showing the different stages of crystal growth for C-S-H at varying pressures. Starting from C+S mixture, crystals formed at lower pressures. At 5 GPa the C-S-H crystals are transparent with transmitted light and by 96 GPa the crystal has metallized and is opaque and reflective. **(b)** Schematic representation of the side-by-side coil system surrounding the diamonds and sample chamber within a DAC. **(c)** Block diagram of the modulated AC susceptibility experimental setup. Excitation and modulation signals are produced by function generators 1 and 2, respectively. In between the brackets are examples of excitation and modulation frequencies. The coil system is installed in a cryostat inside the DAC. The response is measured by a lock-in amplifier in dual reference mode. The basic DAC magnetic susceptibility technique applied here represents an extension of methods applied previously that have led to the discovery and characterization of new superconductors at megabar pressures. The improved method tested on $MgB_2$ and $ErBa_2Cu_3O_7$ demonstrated very clear and distinct peaks at the transition. The superconducting transition obtained with the new instrumentation agrees with the results from the single-frequency method as well as PPMS measurements for both materials used as standards. Using this technique, double-frequency measurements can be conducted on exciting single-frequency coil setups without any modifications to the coils due to the superposition of the excitation and modulation signals. In addition, superposition eliminates the requirement of an extra coil to generate a strong magnetic field to suppress the superconductivity. **(d—f)** Modulated AC magnetic susceptibility measurements of carbonaceous sulfur hydride at several pressures showing the clear peak as the sample enters the superconducting state with maximum $T_c$ of 260 K at 133 GPa. The blue arow indicates the superconducting transition temperature.

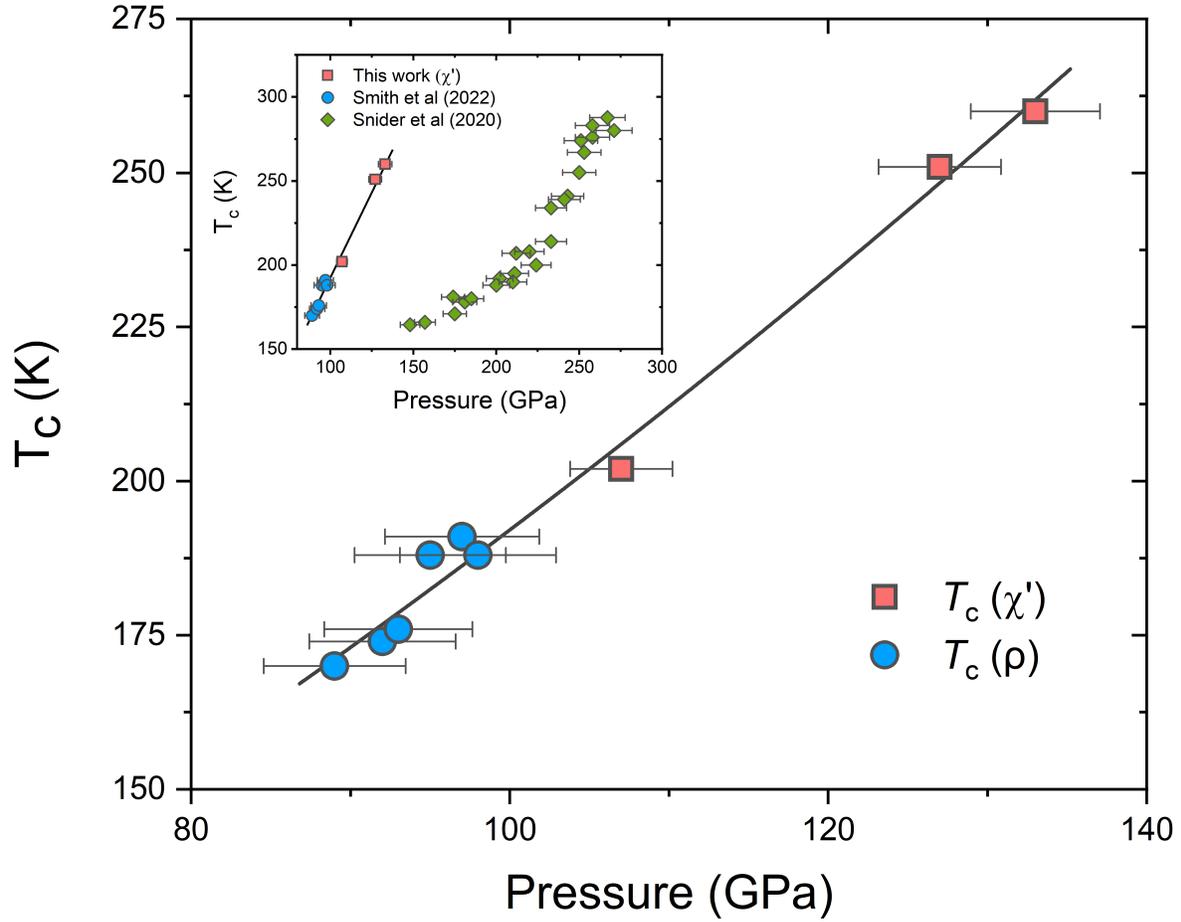

**Figure 2. Pressure dependence of superconducting critical temperature** $T_c$. The modulated AC susceptibility measurements (red squares) reveal a maximum $T_c$ of ~260 K at 133 GPa and follows the response of Smith et. al.,[9] (blue circles) which used electrical resistance measurements to track the transition. The samples in both studies had a similar carbon content but differ from those synthesized and characterized by Snider et al.,[4,5] (green diamonds) and consequently show a different superconducting response, as shown in the inset.

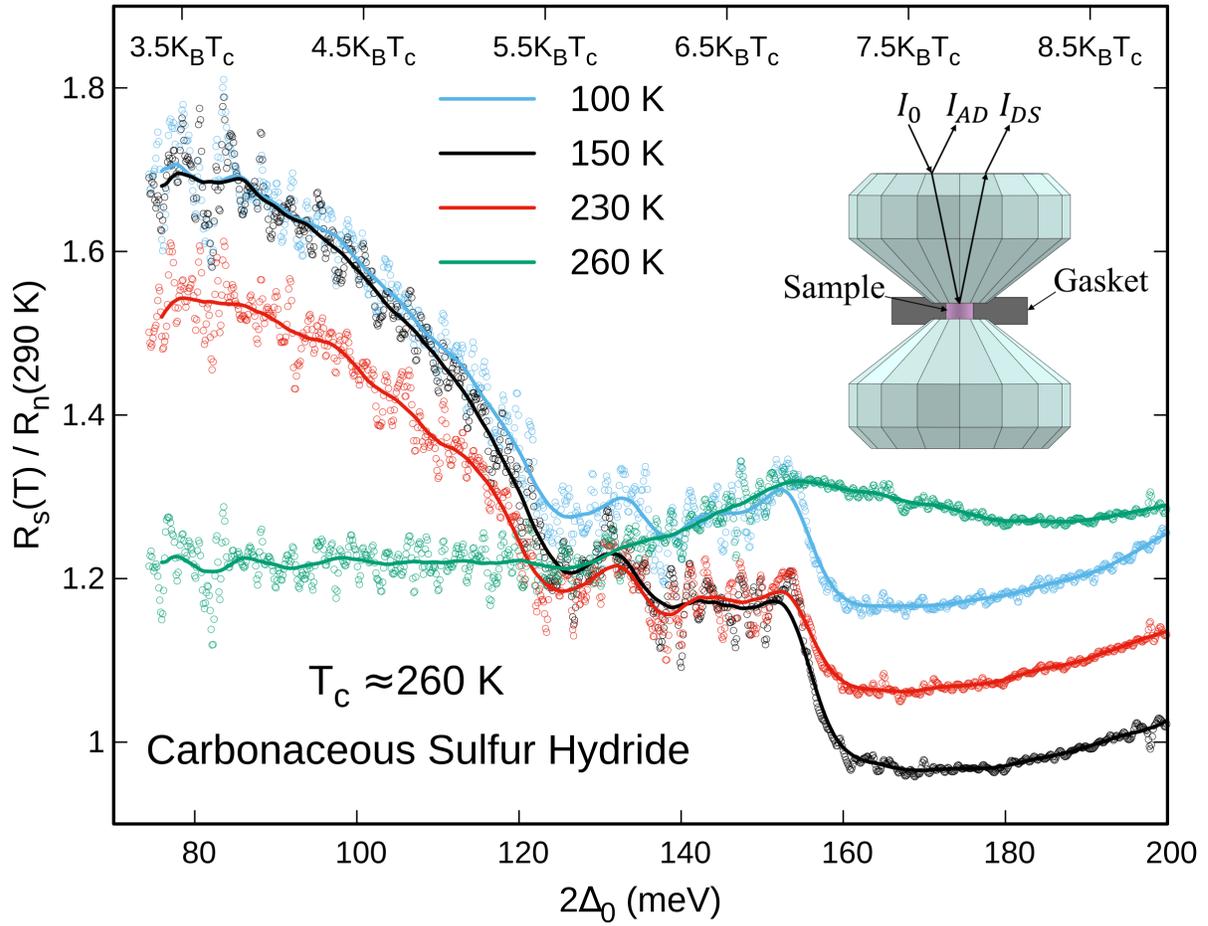

**Figure 3. Synchrotron infrared reflectivity of the C-S-H superconductor.** The $R(T)/R_n$ curves for 260 K, 230 K, 150 K and 100 K, with 290 K (normal state) as the reference. The results give a maximum superconducting gap for the C-S-H sample of ~85 meV at 100 K and 133 GPa. The inset shows the schematic of the ray diagram for reflectivity measurements.

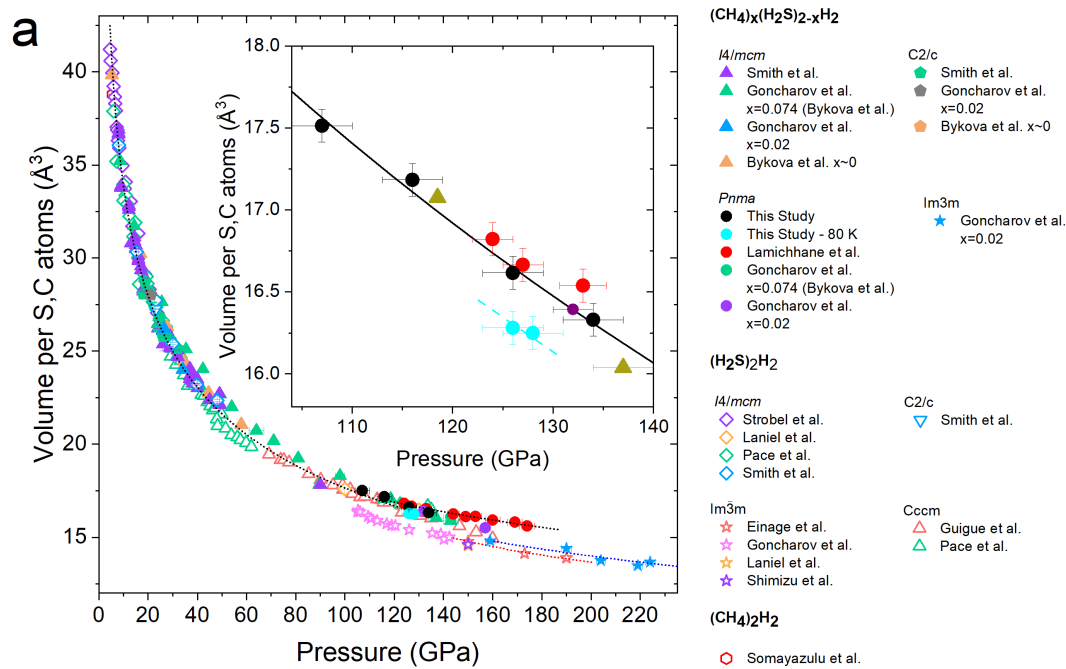

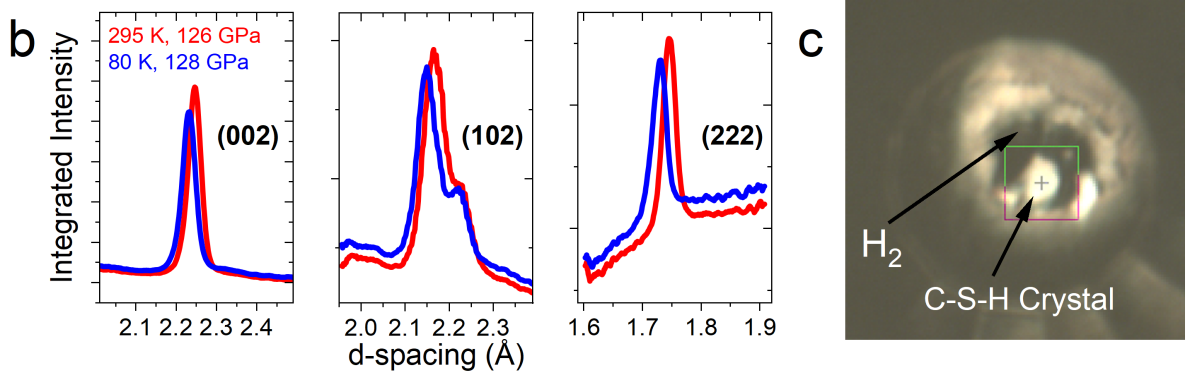
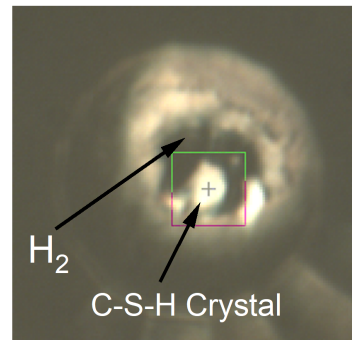

**Figure 4. Synchrotron x-ray diffraction of the C-S-H superconductor.** (a) Pressure-volume relations of C-S-H measured by x-ray diffraction. The black solid circles correspond to the data from this study, and the continuous black line is a Vinet EOS fit to the current C-S-H data with fitting parameters $V_0 = 35(3)$ Å$^3$, $K_0 = 42(12)$ GPa, $K_0' = 4$ (fixed). The light blue solid circles correspond to the data of this study at 80 K and the light blue dash line is guide to the eye. The solid red circles are previous C-S-H data from Lamichhane et al.[55], and the dashed black line is the previous fit to those data. The red open hexagons represent $I4/mcm$ phase of $(CH_4)_2H_2$ points from Somayazulu et. al.[58]; purple open diamonds represents $I4/mcm$ phase of $(H_2S)_2H_2$ from Strobel et al.[59]; green open diamonds and triangles correspond to $I4/mcm$ and $C2/c$ phases of $(H_2S)_2H_2$ respectively from Pace et al.[60]; light red open triangles are $Cccm$ phase of $H_3S$ from Guigue et al.[61]; light purple open stars are $Im\bar{3}m$ phase of $H_3S$ from Shimizu et al.[62]; pink open stars correspond to $Im\bar{3}m$ phase of $H_3S$ reported by Goncharov et al.[63,64]; orange open diamonds and stars represent $I4/mcm$ and $Im\bar{3}m$ phases of $H_3S$ respectively from Laniel el. al.[65]; red open stars are $Im\bar{3}m$ phase of $H_3S$ from Einaga et al.[66]; dotted red line is a fit to the $H_3S$ data from Einaga et al.[66]; light blue open diamonds and inverted triangles represent the $I4/mcm$ and $C2/c$ phases of $(H_2S)H_2$, respectively, from Smith et al.[9]; violet solid triangles and green solid pentagons correspond to the $I4/mcm$ and $C2/c$ phases of $(CH_4)_x(H_2S)_{2-x}H_2$, respectively, where x corresponds to the initial composition of the sample, also from Smith et al.[9]; solid orange triangles and pentagons correspond to the $I4/mcm$ and $C2/c$ phases C-S-H (x~0) data reported by Bykova et. al.[41]; green solid tringles, green solid circles, gray solid pentagons, purple solid circles, and light blue solid stars correspond to $C_xSH$ with x=0.074 $I4/mcm$, x= 0.074 $Pnma$, x=0.02 $C2/c$, x=0.02 $Pnma$, and x=0.02 $Im\bar{3}m$ respectively from Goncharov et. al.[42]; dotted blue line represents fits to the $C_xSH$ data with x= 0.074 $Im\bar{3}m$ phase[42]. (b) Integrated intensity of the (002), (102) and (222) reflections as a function of $d$-spacing describing the Bragg diffraction peaks assigned to the $Pnma$ phase of CSH extracted from the diffraction images in Extended Data Fig. 1 (c) and (d) for 295 K and 80 K at ~ 125(3) GPa. The x-ray was measured in parallel to the AC susceptibility measurements above and below the superconducting transition of ~225 K at ~125(3). (c) Reflection micrograph image of a C-S-H sample in the DAC at 125(3) GPa. The crosshairs identify the highly reflective C-S-H crystals within the sample chamber.

15  Glover, R. E. & Tinkham, M. Conductivity of superconducting films for photon energies between 0.3 and 40k$T_c$. *Phys. Rev.* **108**, 243-256 (1957).

16  Richards, P. L. & Tinkham, M. Far-infrared energy gap measurements in bulk superconducting In, Sn, Hg, Ta, V, Pb, and Nb. *Phys. Rev.* **119**, 575-590 (1960).

17  Bardeen, J., Cooper, L. N. & Schrieffer, J. R. Theory of superconductivity. *Phys. Rev. Lett.* **108**, 1175-1204 (1957).

18  Bardeen, J., Cooper, L. N. & Schrieffer, J. R. Microscopic theory of superconductivity. *Phys. Rev.* **106**, 162-164 (1957).

19  Ashcroft, N. W. Metallic hydrogen: A high-temperature superconductor? *Phys. Rev. Lett.* **21**, 1748-1749 (1968).

20  Ashcroft, N. W. Hydrogen dominant metallic alloys: High temperature superconductors? *Phys. Rev. Lett.* **92**, 187002 (2004).

21  Richardson, C. F. & Ashcroft, N. W. High temperature superconductivity in metallic hydrogen: Electron-electron enhancements. *Phys. Rev. Lett.* **78**, 118-121 (1997).

22  Wang, H., Tse, J. S., Tanaka, K., Iitaka, T. & Ma, Y. Superconductive sodalite-like clathrate calcium hydride at high pressures. *Proc. Natl. Acad. Sci.* **109**, 6463-6466 (2012).

23  Liu, H., Naumov, I. I., Hoffmann, R., Ashcroft, N. W. & Hemley, R. J. Potential high-$T_c$ superconducting lanthanum and yttrium hydrides at high pressure. *Proc. Natl. Acad. Sci.* **114**, 6990-6995 (2017).

24  Peng, F. *et al.* Hydrogen clathrate structures in rare earth hydrides at high pressures: Possible route to room-temperature superconductivity. *Phys. Rev. Lett.* **119**, 107001 (2017).

25  Bi, T., Zarifi, N., Terpstra, T. & Zurek, E. in *Reference Module in Chemistry, Molecular Sciences and Chemical Engineering*   (Elsevier, 2019).

26  Zurek, E. & Bi, T. High-temperature superconductivity in alkaline and rare earth polyhydrides at high pressure: A theoretical perspective. *J. Chem. Phys.* **150**, 050901 (2019).

27  Semenok, D. V., Kruglov, I. A., Savkin, I. A., Kvashnin, A. G. & Oganov, A. R. On distribution of superconductivity in metal hydrides. *Curr. Opin. Solid State Mater. Sci.* **24**, 100808 (2020).
23

## Methods

**Sample preparation.** A number of high pressure experiments were carried out with consistent reproducibility of Raman, electric resistance data, magnetic susceptibility studies. The samples were loaded onto a membrane-driven diamond anvil cell (m-DAC), using 1/3-carat, type II diamond anvils with a 0.2, 0.15, 0.1, and 0.05 (for higher pressures) mm culet with bevels up to 0.3 mm at 8°. A 0.25 mm thick rhenium gasket was pre-indented to 8 to 20 μm (depending on the pressure) and a 120 (or 70, 30 for high pressures) μm hole was electro-spark drilled at the center of the gasket. The mixture is ball-milled to a particle size less than 5 μm. We loaded a clump of a C + S powder mixture into a diamond anvil cell (DAC), after which molecular hydrogen is loaded to serve as both reactant and pressure transmitting medium (PTM). Raman scattering confirmed the presence of the starting materials in the DAC. The confirmed DAC samples were compressed to 4.0 GPa and exposed to 532 nm laser light for several hours at a power of 10-25 mW to synthesize the C-S-H crystals that can grow as big as 80 μm in diameter.

**Hydrogen loading and Raman spectroscopy.** Highly compressed gases for loading into the m-DAC are necessary in order to obtain sufficient density. The density of gases at room temperature and atmospheric pressure is too low to obtain a sufficient quantity of sample after they condense into fluid or solid phases. The higher density provides enough hydrogen after collapsing the sample chamber as the pressure increases. We thus used a high-pressure gas loader to compress gases to high densities. The m-DACs were first loosely closed and mounted into a gearbox. Under a microscope, the DACs were opened 90 degrees using the central gear of the gearbox. The DAC and gearbox were then placed into a high-pressure gas loader (Top Industries). The system was first flushed with hydrogen to purge the circuit of impurities. The sample chamber was pressurized to ~2500-3000 bar. The gas loader was then drained, and the DAC was retrieved from the gearbox.

Also, in some experiments, high purity hydrogen gas was cryogenically loaded into a DAC in a cryostat mounted on an optical table with $CaF_2$ IR transmitting windows. In order to contain the liquid $H_2$ we have used a mini chamber with a balseal[67-69]. A capillary tube was attached to the mini chamber for the gas to flow. The optical table also incorporated standard instrumentation to measure Raman scattering and fluorescence.

Raman spectra were collected using a custom micro-Raman setup using 532 nm Millenia eV laser, Princeton Instruments HRS 500 spectrometer, and a Pylon camera. The laser was focused using a 20× zoom objective G Plan Apo 20× Objective. The backscattered light travels through a spatial filter and volume Bragg-based notch filters allowing acquisition of low wavenumber Raman peaks. The sample, laser spot, and spatial filter were imaged and aligned using a mirror on a flip stand directing the light to a PointGrey CCD camera.

**Modulated AC magnetic susceptibility.** To perform a measurement, an SR860 lock-in is used in Dual Reference mode. The lock-in is set to internal reference at the second harmonic of the modulation frequency, and external reference of the excitation frequency. The Dual Reference mode is thus able to measure the modulation of the excitation response at the second harmonic of the modulation - *i.e.,* the sample being driven in and out of the superconducting state. This manifests itself as an asymmetric peak in voltage.

The coil system used in the double-frequency method included two identically wound pairs of coils, prepared using a 46 AWG copper wire. The first pair consisted of a primary drive coil and a secondary pick-up coil each with 160 turns. The other (dummy primary and dummy secondary) pair is identical to the first pair. The primary coils from the two pairs were connected in series in the opposite direction. Secondary coils were connected in series as well, but with same polarities. The primary coils create an alternating magnetic field, which induces an electromotive force on the secondary coils. Since primary coils were wired with opposite polarity, the induced electromotive force cancels each other and give zero net electromotive force. However, in practice, the cancellation may not be complete, and the coils can be balanced to reduce the background by removing turns from one of the primary coils.

The sample and the gasket were placed inside the first set of coil pairs. To maximize the signal strength, the sample was positioned completely inside the coils, which increases the magnetic field lines passing through it. Also, the coils were much closer and in the same plane with the sample, a small coil with fewer turns create the necessary stronger magnetic field. This reduces the size of the coil setup and can be adapted to use with most DAC designs. Furthermore, to reduce the magnetic flux deformation, nonmagnetic DACs and cubic boron nitride (cBN) anvil seats were used.

The high-frequency, low-amplitude excitation signal and the low-frequency, high-amplitude modulation signal were generated by two function generators. Then low- and high frequency signals are combined and fed to a voltage-to-current converter. This will convert the voltage signal into a current waveform. Then this signal is fed into the primary coils. During tests with $MgB_2$ and $ErBa_2Cu3O_7$ (see Extended Data figure 5), it was found that typically 500 - 2000 Hz (driven at 5-10 mA) for the excitation signal and 20-250 Hz (driven at 90-100 mA) for the modulation signal generate the best results.

Signal (induced voltage) from the secondary pickup coils first goes through a pre-amplifier and amplifies to a suitable range for the lock-in amplifier input. Conventionally, the double-frequency method signal was captured by using two lock-in amplifiers. First, it takes the signal from the pickup coil and demodulates at $F_{exc}$. The output is then fed to another lock-in to demodulate the $F_{mod}$ signal. With this method, the time constant of the first lock-in needs to be carefully adjusted to pass the modulated signal to the second lock-in. To facilitate this synchronization we used a lock-in amplifier cable with a dual reference mode (*i.e.,* SRS 860/SRS 865). The excitation frequency was used as the external reference for the lock-in, while the excitation second harmonic of the modulation frequency is used as the internal reference. In this mode lock-in directly demodulates the sidebands at $|F_{exc} - 2 * F_{mod}|$ by amplitude modulation. This method thus does not require an additional lock-in amplifier, which simplifies the experiment. The approach also leads to improved SNR compared to previous methods and is less sensitive to the lock-in settings. A small temperature dependent background signal is observed, but the transition is clearly visible despite the background. In this work, cubic polynomial backgrounds were used as a background.

**Synchrotron infrared spectroscopy.** Synchrotron-based IR reflectivity was measured on the same DAC samples at beamline 22-IR-1 of the National Synchrotron Light Source II (NSLS-II) at Brookhaven National Laboratory (BNL). The synchrotron beam extracted from the large-gap IR dipole of the NSLS-II storage ring provides extremely stable IR source through a top-off operating mode (399-401 mA beam variation). The optical system (EDF. 3) includes a Bruker Vertex 80 FT-IR spectrometer coupled to a custom long working-distance IR microscope with a MCT detector and KBr beamsplitter. The procedure used to analyze the spectra is summarized as follows. Consider a diamond anvil cell with a sample under pressure and assume all electromagnetic rays are at normal incidence at corresponding interfaces.

First consider the incident light at the diamond anvil table.

$$I_{AD} = I_0 R_{AD} \kappa \qquad (1)$$

where $I_0$ is the incident intensity, $I_{AD}$ is intensity counts from air-diamond interface, $R_{AD}$ is reflectivity at air-diamond interface and $\kappa$ is optical loss in the system[70]. Next consider reflected light from the sample inside the DAC, where $I_{DS}$ is the intensity counts from the sample inside the chamber.

$$I_{DS} = I_0(T_{AD}\eta(\lambda))(R_{DS}\eta(\lambda))T_{AD}\kappa \qquad (2)$$

where $T_{AD}$ is the transmittance at air-diamond interface, $R_{DS}$ is reflectivity at diamond-sample interface and $\eta(\lambda)$ is diamond absorption as a function of wavelength. By combining the two equations, reflectivity of the sample is calculated as follows,

$$R_{DS} = \left(\frac{I_{DS}}{I_{AD}}\right)\left(\frac{R_{AD}}{(1-R_{AD})^2}\right)\frac{1}{\eta(\lambda)^2} \qquad (3)$$

Assuming the diamond refractive index as 2.42, $R_{DS}$ can be simplified to

$$R_{DS} = 0.25\left(\frac{I_{DS}}{I_{AD}}\right)\frac{1}{\eta(\lambda)^2} \qquad (4)$$

In static high-pressure environments, measuring absolute reflectivity is challenging. Calculation of that quantity requires diamond absorption which is unknown and a unit reflectance surface inside the DAC which does not exist. In this work, instead of calculating the absolute reflectivity of the sample, we compare the measured IR reflected intensities in both the normal and superconducting state. The resulting relative reflectivity is then calculated as the ratio of, $R_s/R_n$. Where $R_s$ is reflectivity of the superconducting state and $R_n$ is the normal state reflectance of the sample. Similar analyses have been carried out for other superconducting materials including sulfur hydride[48,71-73]. Using Equation 4, $R_s/R_n$ is simplified to,

$$\frac{R_s}{R_n} = \frac{I_s}{I_n} \quad (5)$$

where $I_s$ is intensity counts from superconducting state when $T < T_C$ and $I_n$ is intensity counts from normal state when $T > T_C$.

**Synchrotron x-ray diffraction**. Synchrotron x-ray diffraction measurements were carried at three different beamlines HPCAT (Sector 16) beamlines 16-ID-D, 16-BM-D, and GSECARS (Sector 13) beamline 13-ID-D at the Advanced Photon Source, Argonne National Laboratory. *In-situ* high-pressure low-temperature diffraction measurements were carried simultaneously with magnetic susceptibility above and below $T_c$ at beamline 16-ID-D. A specially designed DAC for magnetic susceptibility measurement mounted in a custom-built flow type l-$N_2$ cryostat. The cryostat is designed for a stable sample holder for zero sample motion during cryogenic cooling of a DAC and a double membrane compression/ decompression drive, for precise pressure control. The modular flexible design and abundance of electrical feedthroughs on the cryostat allowed the integration of multiple analytical techniques, *e.g.,* magnetic field modulation, susceptibility, temperature control, etc. The incident radiation of wavelength 0.4066 Å was focused to a spot size of ~6 × 8 μm² (FWHM) at the sample position. Angle-dispersive diffraction patterns were recorded at high pressure of 107(3) and 125(3) GPa and from ambient temperature to up to l-$N_2$ temperature during the multiple heating and cooling cycles. As the DAC used was designed for observations of superconductivity, its opening angle for diffraction was only ±5°, which reduced the reciprocal space available for crystal structure refinement to identify the atomic positions. Diffraction patterns were collected at 116(3) GPa and 133(3) GPa at ambient temperature at beamlines 13-ID-D and 16-BM-D, with wavelengths 0.3344 Å and 0.4133 Å and focused spot sizes of ~3 × 4 µm$^2$ (FWHM) and ~8 × 5 µm$^2$ (FWHM), respectively. The pressure for each diffraction measurement was determined from diffraction of the Re gasket adjacent to the sample, as confirmed by measurements of diamond anvil Raman edge measurements at selected pressures. At 16-ID-D beamline x-ray images were recorded on a PILATUS 1M Silicon detector and at 16-BM-D and 13-ID-D beamlines x-ray images were recorded on a PILATUS CdTe 1M detector. X-ray images were integrated using DIOPTAS software.

# Supplemental Information

**Table S1.** Experimental lattice parameters and unit-cell volumes of single-crystal C-S-H sample as a function of pressure and temperature.

| P (GPa) | T (K) | $a$ (Å) | $b$ (Å) | $c$ (Å) | V/S(C) (Å$^3$) |
|---|---|---|---|---|---|
| 107(3) | 295 K | 7.86(2) | 7.75(2) | 4.60(2) | 17.51(10) |
| 116(3) | 295 K | 7.82(2) | 7.71(2) | 4.56(2) | 17.18(10) |
| 126(3) | 295 K | 7.75(2) | 7.64(2) | 4.49(2) | 16.62(10) |
| 133(3) | 295 K | 7.71(2) | 7.60(2) | 4.46(2) | 16.33(10) |
| 126(3) | 80 K | 7.71(2) | 7.59(2) | 4.45(2) | 16.28(10) |
| 128(3) | 80 K | 7.70(2) | 7.59(2) | 4.45(2) | 16.25(10) |

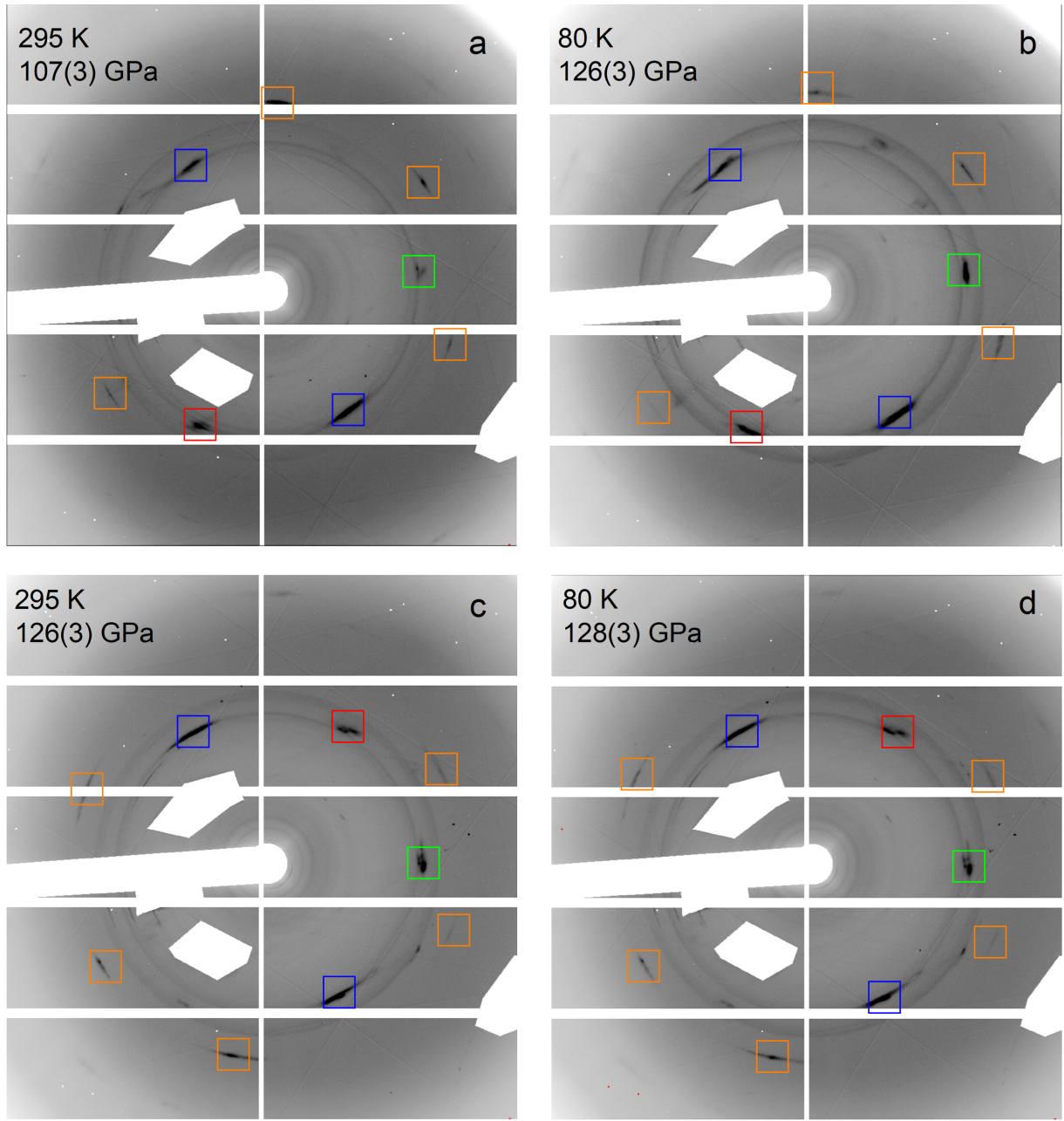

**Extended Data Figure 1**. Comparison of single crystal diffraction images of the C-S-H sample obtained at 16-ID-D HPCAT beamline at (a) 295 K and 107(3) GPa and (b) 80 K and 126(3) GPa. (c) 295 K and 126(3) GPa and (d) 80 K and 128(3) GPa. There is a ~180 rotation of the x-ray

image in (c) and (d) because the DAC was remounted in the cryostat following diamond anvil Raman edge pressure measurements outside the cryostat. The main reflections indexed based on *Pnma* symmetry are shown. 002 planes (blue boxes), 222 planes (orange boxes), 012 planes (green box) and 102 planes (red box). The two complete rings correspond to diffraction from polycrystalline Re and ReH gasket material. White regions from center to the edge are masked out shadows of the x-ray beam, and the white crossed lines arise from gaps in the detector.

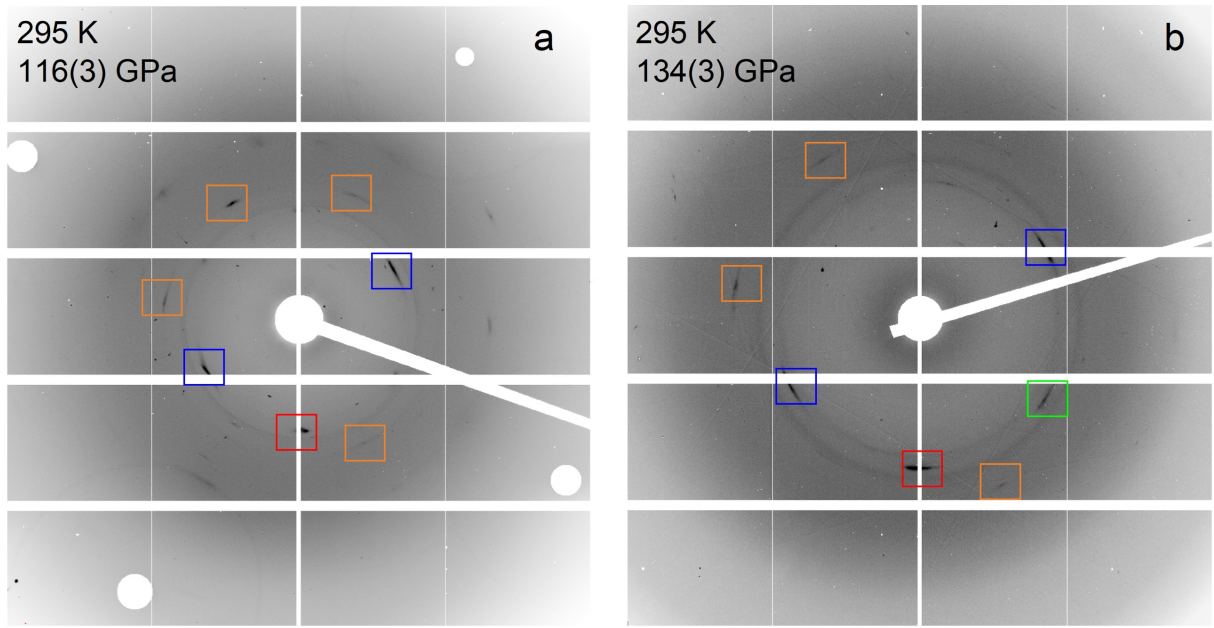

**Extended Data Figure 2.** Selected single crystal diffraction images of the C-S-H sample obtained 295 K. (a) using GSECARS 13-ID-D 295 K at 116(3) GPa and (b) using HPCAT 16-BM-D beamline at 134(3) GPa. The main reflections indexed using *Pnma* symmetry are shown. 002 planes (blue boxes), 222 planes (orange boxes), 012 planes (green box) and 102 planes (red box). The two complete rings are diffraction from the polycrystalline Re and ReH gasket material.

**Extended Data Figure 3.** Schematic of the optical set up specifically configured for the synchrotron-based IR reflectivity measurements.

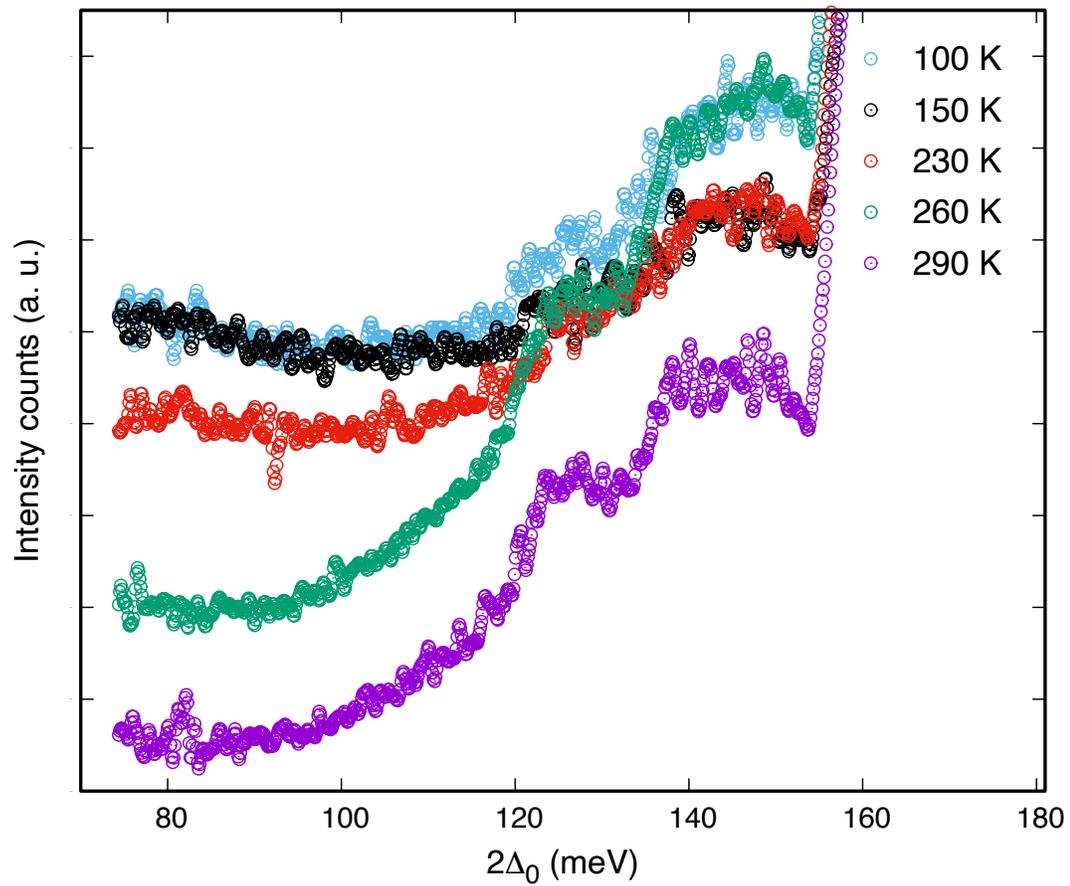

**Extended Data Figure 4.** The raw IR reflectance measurements from a C-S-H sample at 100 K, 150 K, 230 K, 260 K and 290 K and 133 GPa at 70-170 meV. As the sample cools from 290 K to 260 K.

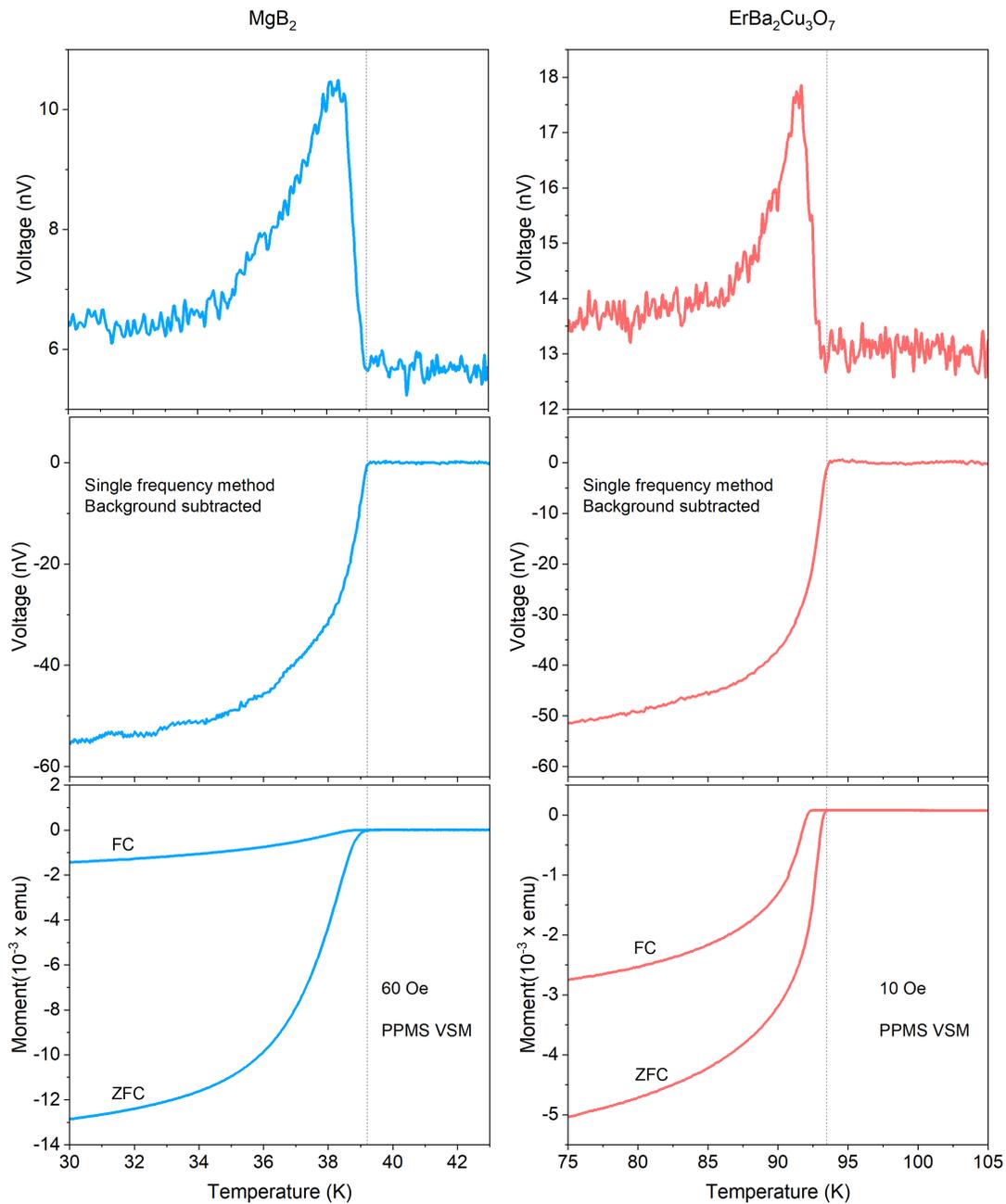

**Extended Data Figure 5.** *Top left*: The superconducting signal of MgB$_2$ sample using frequency modulation method with 1.42 kHz excitation signal and 98 Hz modulation signal. Top right: Superconducting signal of ErBa$_2$Cu$_3$O$_7$ using same method with 1.9 kHz excitation signal and 147 Hz modulation signal. Signal is divided by pre-amplifier factor (x100 for MgB$_2$ and x500 for ErBa2Cu3O7) for comparison. *Middle left*: Superconducting transition of MgB$_2$ with the same sample and same coil system with the single frequency method. *Middle right*: Superconducting transition of ErBa$_2$Cu$_3$O$_7$ with the same sample and same coil system with the single frequency method. *Bottom left:* Superconducting transition using a standard instrument (PPMS), showing field (FC) and zero field cooling (ZFC) of MgB$_2$. Bottom right: Superconducting transition using a standard instrument (PPMS) measurements with field (FC) and zero field cooling (ZFC) of ErBa2Cu3O7.